\documentclass[12pt,preprint]{aastex}
\usepackage{natbib}

\def\lesssim{\mathrel{\hbox{\rlap{\hbox{\lower4pt\hbox{$\sim$}}}\hbox{$<$}}}}
\def\gtrsim{\mathrel{\hbox{\rlap{\hbox{\lower4pt\hbox{$\sim$}}}\hbox{$>$}}}}

\def\micron{\hbox{$\mu$m}}
\def\arcdeg{\hbox{$^\circ$ }}

\def\arcsec{\hbox{$^{\prime\prime}$ }}

\begin{document}

\title{The First Spatially Resolved Mid-IR Spectra of NGC 1068 Obtained at 
Diffraction-limited Resolution with LWS at Keck I Telescope}
\author{JOSEPH H. RHEE}
\email{jrhee@gemini.edu}
\affil{Gemini Observatory, 670 N. A'ohoku Place, Hilo, HI 96720}
\author{JAMES E. LARKIN}
\email{larkin@astro.ucla.edu}
\affil{Dept. of Physics and Astronomy, University of California, Los Angeles}

\begin{abstract}
We present spatially resolved mid-IR spectra of NGC 1068 with a 
diffraction-limited resolution of 0.25\arcsec using the Long Wavelength 
Spectrometer (LWS) at the Keck I telescope.  The mid-infrared image of NGC~1068 
is extended along the N-S direction.  Previous imaging studies have shown the 
extended regions are located inside the ionization cones indicating that the 
mid-infrared emission arises perhaps from the inner regions of the narrow-line 
clouds instead of the proposed dusty torus itself.  The spatially resolved 
mid-IR spectra were obtained at two different slit position angles, +8.0 and 
-13.0 degrees across the elongated regions in the mid-IR.  From these spectra, 
we found only weak silicate absorption toward the northern extended regions but 
strong in the nucleus and the southern extended regions.  This is consistent 
with a model of a slightly inclined cold obscuring torus which covers much of 
the southern regions but is behind the northern extension.  While a detailed 
analysis of the spectra requires a radiative transfer model, the lack of 
silicate emission from the northern extended regions prompts us to consider a 
dual dust population model as one of the possible explanations in which a 
different dust population exists in the ionization cones compared to that 
in the dusty torus.  Dust inside the ionization cones may lack small silicate 
grains giving rise to only a featureless continuum in the northern extended 
regions while dust in the dusty torus has plenty of small silicate grains to 
produce the strong silicate absorption lines towards the nucleus and the 
southern extended regions.
\end{abstract}


\section{Introduction}
AGNs appear in a variety of types, which are often classified based on the 
presence of broad emission lines.  Objects are classified as type I if both 
broad and narrow emission lines appear in their optical spectra and as type 
II if only  narrow emision lines are present without broad emission lines.  
It is, however, a general belief that much of the observed diversity in the 
local universe arises from different viewing angles toward the central engine 
and a dusty toroidal structure around it, especially for Seyfert galaxies of 
types I and II.  When the dusty torus is viewed face-on, both the central 
engine and the broad-line regions can be seen directly causing objects to 
appear as Seyfert 1 galaxies.  When the dusty torus is viewed edge-on, the 
anisotropic obscuration created by the torus causes objects to appear as 
Seyfert 2 galaxies (see \citealt{ant93} for review).  It is this crucial 
role played by dust in the Unified model of AGN that makes understanding 
dust properties very important in understanding AGN.  

A significant fraction of the optical/UV/X-ray luminosity of the active 
nucleus is absorbed by the proposed dusty torus and reradiated at mid-infrared 
wavelengths.  The infrared also suffers less extinction than the optical band 
and is prefered for probing the proposed dusty torus or dust in general in the 
nucleus.  Early mid-infrared observations of Seyfert galaxies \citep{roc91} 
have shown that the 9.7 $\micron$ silicate feature appears in strong absorption 
in Seyfert 2's as expected from an edge-on geometry of the proposed dusty torus 
for type II objects.  If the dust responsible for the 
9.7 $\micron$ silicate feature belongs to the dusty torus, then the spatial 
distribution of the silicate absorption can provide very important clues to the 
physical properties (size, orientation, etc.) of the dusty torus.  

It is, however, difficult to investigate the spatial distribution of dust in 
Seyfert galaxies because mid-IR emission from most Seyfert galaxies has not been 
resolved (see \citealt{gor04}).  While thermal emission from hot dust has been 
considered as a dominant source of the mid-IR emission of Seyfert galaxies, the 
unresolved nature of the mid-IR image have left the discussion over a non-thermal 
origin still alive.  NGC~1068 is one of a few Seyfert galaxies whose mid-IR 
emission is resolved.  No other alternative mechanism like synchrotron radiation 
can produce emission over such an extended area, leaving heated dust grains as 
the likely source.  

NGC 1068 is classified as a Seyfert 2 based on the presence of narrow 
emission lines and absence of broad emission lines.  The detection of 
broad emission lines in polarized light \citep{ant85}, however, has shown 
that NGC 1068 harbors an obscured Seyfert 1 nucleus.  As one of the closest 
and brightest Seyfert 2 galaxies, NGC 1068 offers a better spatial scale for 
the detail investigatation of its obscured nucleus.  NGC 1068 is at 14.4 Mpc 
\citep{tul98} so 1\arcsec corresponds to 72 pc or 0.25\arcsec (our spatial 
resolution) corresponds to 18 pc in physical distance.  As a part of a greater 
mid-infrared survery of a sample of Seyfert galaxies, NGC 1068 was selected 
specifically for the investigation of the physical properties of the proposed 
dusty torus.

NGC 1068 has been observed many times in the mid-infrared both in imaging and 
spectroscopic modes. The spatially resolved mid-IR images of NGC~1068 show 
a linear structure covering about 1\arcsec in the north-south direction 
\citep{bra93,cam93,boc00,tom01}.  Emission line imaging of O[III] showed that this 
disklike structure lies inside the narrow-line regions \citep{eva91,mac94}.  
Since the narrow-line regions are believed to be created by the ionizing 
radiation from the central engine which is collimated by the dust torus, 
the dusty torus should be oriented in the East-West direction perpendicular 
to the narrow-line cones.  Thus this structure has been believed to be 
created by grains in the NLRs heated by the nuclear radiation.  In this 
model, the dusty torus is then too cold to emit significantly at 10 \micron.

While recent mid-infrared imaging studies have provided some spatially resolved 
dust measurements in the nuclear region of NGC 1068, no spatially resolved 
spectra have been obtained in the mid-infrared until this study.  Mid-infrared 
spectra of NGC 1068 have also been obtained repeatly by several single aperture 
ground-based telescopes and Infrared Space Observatory (ISO) using various 
apertures from 0.4\arcsec at VLTI to 24\arcsec $\times$ 24\arcsec at ISO 
\citep{lut00,lef01,jaf04,roc91,sie04}.  Most previous spectra show that the 
mid-infrared spectra of NGC 1068 have significant silicate absorption.
Especially \citet{stu00} and \citet{lut00} show the AGN dominated 
spectra with no significant PAH emission lines.

A crude spatial map of the distribution of the silicate absorption has been 
provided by imaging studies using multiple narrow bands, most notably from 
\citet{boc00} and \citet{gal03}.  However, these studies have given conflicting 
results making spatially resolved spectra necessary to resolve the issues.  
For example, \citet{boc00} has reported that the silicate feature is 
relatively strong in absorption on the nucleus and to the south but flat or 
even in emission to the north.  In contrast, \citet{gal03} found that the 
silicate feature appears strong in absorption to the north but either flatter 
or in emission to the south.  For the first time, we present the 
diffraction-limited mid-infrared spectra of NGC 1068 at a spatial resolution of 
0.25\arcsec to investigate the spatial distribution and the properties of dust 
in the nuclear regions.


\section{Observation \& Data Reduction}
The mid-infrared spectra of NGC~1068 were obtained on September 6th, 2003 as 
a part of a larger mid-infrared spectroscopic survey of nearby Seyfert galaxies.  
The low-resolution spectroscopic mode of the Long Wavelength Spectrograph 
(LWS; \citealt{jon93}) was used at the f/25 forward Cassegrain focus of the 
Keck I 10 m telescope.  LWS uses a 128 x 128 pixel Boeing Si:As array with a 
spatial pixel scale of 0.08\arcsec.  For the present spectra, a slit with a 
3 pixel width was used in the bottom 3.6\arcsec (45 pixels) of the array giving 
a 0.24\arcsec by 3.6\arcsec of slit size.  The spectral dispersion is 0.0375 
$\micron$ per pixel.  The N-wide filter with a central wavelength of 10.1 
$\micron$ was selected to give wavelength coverage of 7.71 - 12.48 $\micron$.  
This wavelength range includes prominent PAH emission lines at 8.6 $\micron$ 
and 11.3 $\micron$ as well as a wide silicate absortion line at 9.7 $\micron$. 

A log of our observations is presented in Table 1.  The observations were made 
under excellent atmospheric conditions with low water vapor.  Air masses varied 
between 1.06 and 1.11.  The telluric absorption was measured by observing 
several standard stars, HR 337, HR 617, HR 8781, and HR 1017. 
The spectra of NGC 1068 were obtained at three different 
slit position angles, +8.0, +78.9, and -13.0 degrees which were selected based 
on both the direction of the extended mid-infrared core and the location of 
bright guide stars.  In LWS, the slit position angle on the sky is determined 
by the location of a bright offset guide star, which controls the telescope 
tracking.  The guide star for the slit position angles of +8.0$^{o}$ 
and -13.0$^{o}$ were bright enough to give stable tracking.  But the guide 
star at +78.9$^{o}$ was not bright enough so tip-tilting errors were visible.

The standard ``chop-nod'' mode was used in the observation in order to suppress
sky emission and radiation from the telescope.  The field was chopped at a 
frequency of 2 Hz and nodded every 30 sec.  The chopping and nod directions were 
set to the same direction, parallel to the slit.  The nodding amplitude was 
the same as the chopping amplitude of 10$\arcsec$.  An individual spectrum set 
was created by coadding a total of 60 frames with 50 ms integration time at 
each chop beam.  These individual sets were then combined to produce 7 nodsets 
per each runtime with total on-source integration time of 168 sec.  Three 
runtime spectra were acquired at each slit position angle for NGC~1068.

The raw spectra for both NGC~1068 and the standard stars were first sky-subtracted using 
the sky frame of the chop pair and flat-fielded.  Bad pixels were replaced by 
the median of their neighboring pixels.  The resultant spectra were then 
spatially rectified by a first order polynomial before shifting the images 
parellel to the slit to align the peak position and in the direction of 
dispersion to align atmospheric absorption lines.   After combining the images 
from each runtime set, a wavelength calibration was applied to them using 
atmopheric transmission lines\footnote{The actual data set of the atmospheric 
transmission lines were obtained from the website of the Gemini Observatory.} 
with wavelengths calculated by \citet{lor92}.   

\begin{figure}
\figurenum{1}
\epsscale{.90}
\plotone{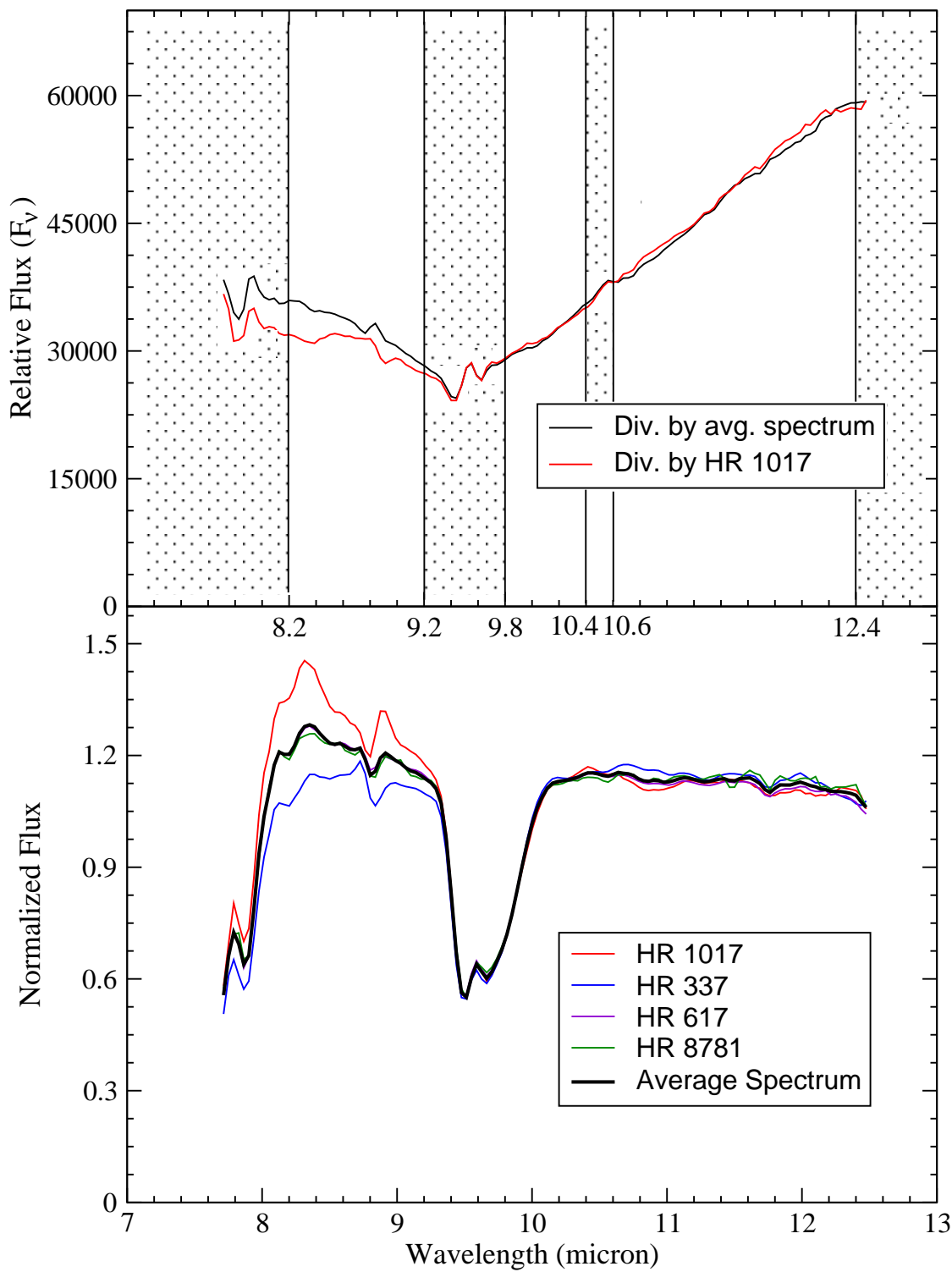}
\caption{Calibration of the mid-IR spectra of NGC 1068.  The top panel shows 
two mid-IR spectra of NGC 1068 at the nucleus: one (black line) divided 
by the average spectrum of 4 standard stars and the other (red line) 
divided by the spectrum of HR 1017.  The bottom panel displays specta of 
all 4 standard stars.  When divided by the spectrum of HR 1017, the mid-IR 
spectrum of NGC 1068 produces unknown broad emission-like features between 
8.4 and 8.8 microns.  The shaded regions in the top panel indicate the 
wavelength regions omitted from the fitting process.}
\label{n1068_cal4}
\end{figure}

The average spectrum of 4 standard stars was divided by a simulated blackbody 
spectrum with the star's effective temperature and then divided into each 
runtime spectrum of NGC~1068 in order to remove atmospheric and instrumental 
features including the deep ozone absorption band at 9.6 $\micron$.  
HR 1017, a F5 star, was observed closest in time to NGC 1068 but at a 
worse airmass, 1.17.  After many trials, we found that the average spectrum 
of all four stardard stars yielded the best telluric correction in the spectra of 
NGC 1068.  The upper panel of Figure \ref{n1068_cal4} shows the mid-infrared 
spectra of NGC 1068 at the nucleus divided by the spectrum of either HR 1017 
or the average of 4 standard stars.  The bottom panel displays the average 
spectrum of all 4 standard stars as well as their individual spectrum.  When 
divided by the average spectrum, the spectrum of NGC 1068 looks much smoother 
in the short wavelength regions between 8.0$\micron$ and 9.3$\micron$.  

Before extracting the 
one-dimensional spectra of NGC~1068, 14 individual chop sets in each runtime 
set were inspected in order to check whether the seeing was stable throughout 
the exposure.  Although three runtime sets were acquired at each position angle, 
only one runtime set was found to be stable at the slit position angles of 
+8.0$^{o}$ and -13.0$^{o}$.  Thus one-dimensional spatially-resolved spectra 
were obtained by extracting consecutive 3 rows (0.25$\arcsec$) in the spatial 
direction from the maximum emission of the single stable runtime set. 
For the total slit spectra, total 24 rows (2\arcsec) were combined to produce 
1-D spectra from the average of 3 runtime sets.

Once the 1-D spectra were extracted for each galaxy, we used a foreground 
screen model consisting 
of a power law continuum and silicate absorption line to fit the observed 
silicate feature.  The silicate extinction curve was produced by \citet{dud97} 
using the $\mu$ Cep emissivity curve given by \citet{roc84}.  We used four free 
parameters for the fitting: the amplitude of the power law; a power-law index, 
s, for $F_{\lambda}$ $\sim$ $\lambda^{s}$; a visual extinction, $A_{v}$; and 
the central wavelength, $\lambda_{c}$.  We excluded the wavelength regions of 
strong atmospheric contamination and [SIV] emission ($\lambda$ = 10.511 $\micron $ 
in the rest frame) and came up with the three wavelength intervals, 8.2 - 9.2 
$\micron$, 9.8 - 10.4, 10.6 - 12.4 $\micron$ that were used for the fit 
(Figure 1).  The silicate absorption profile was redshifted to match the 
observed spectra before the fitting process.  We began with 
a very rough power-law fit to estimate the initial amplitude and the slope.  We 
also get a rough estimate of the visual extinction and the central wavelength 
by eye.  Then we ran a grid search to find a best fit to the observed spectrum.  
First, we started the search within $\pm$ 40 per cent of the initial values for 
the first 3 parameters in 1 per cent increments.  For the central wavelength, 
the iteration was run within $\pm$ 2 $\micron$ from the best guessed centeral 
wavelength in 0.1 $\micron$ increment.  Then from the best fit result of the 
initial search, we ran the grid search again with a 0.1 percent increment 
within $\pm$ 20 per cent of the revised center.  Most times the fit converged 
but sometimes it did not.  Once the best fit was obtained, we measured the 
calculated Chi Square ($\chi^{2}$/N) using the same three intervals that we 
used for the fitting to obtain the standard deviation of our spectra.  The 
noise of our mid-infrared data is background limited and does not vary much 
outside the spectral region contaminated by atmopheric absorption.  Three 
spectra with $\chi^{2}$/N under unity suggest that the reduced Chi Square 
value is perhaps underestimated by an overestimation of the noise in the spectra. 


\section{Results}

\begin{figure}
\figurenum{2}
\epsscale{.75}
\plotone{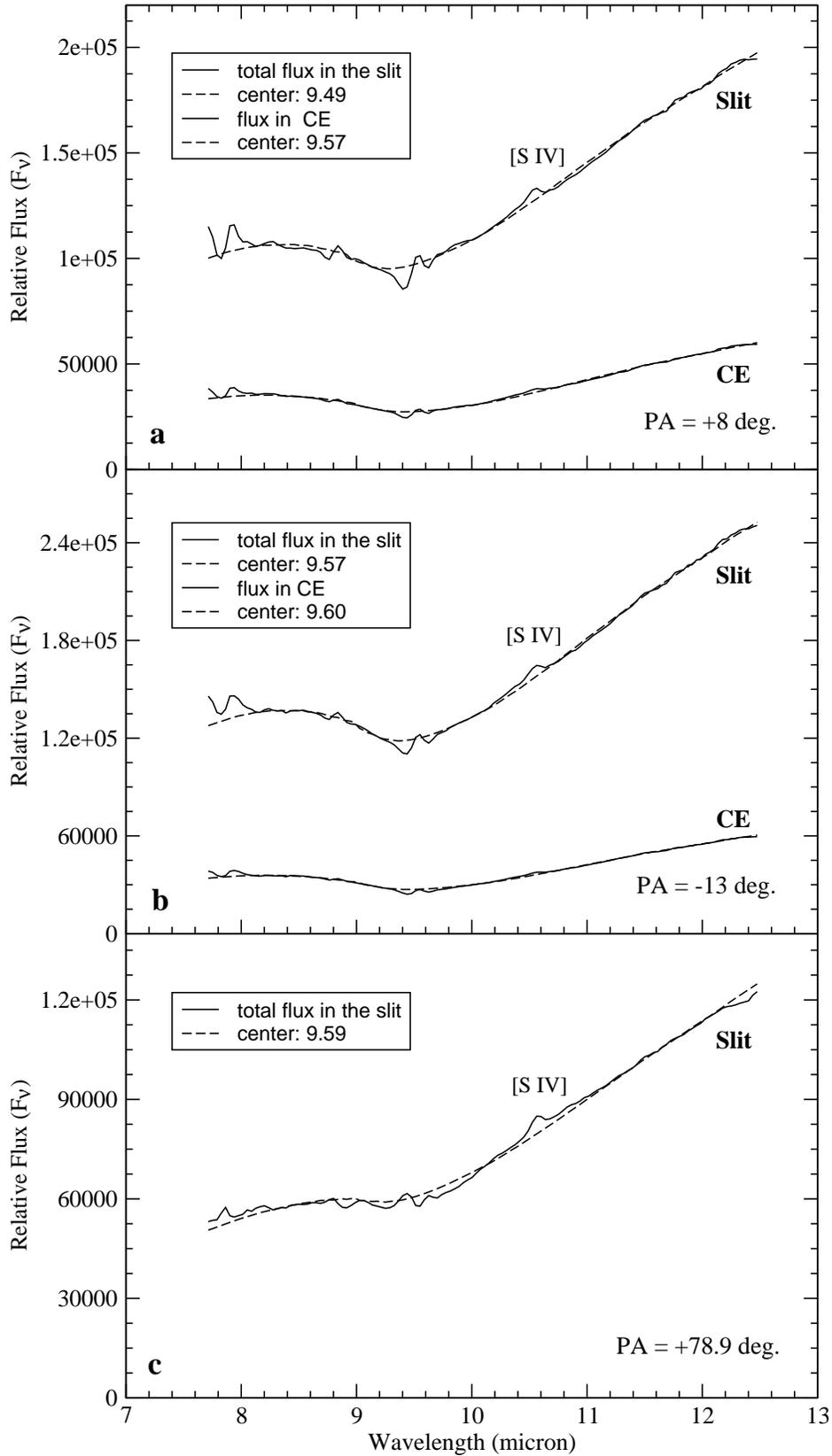}
\caption{Mid-infrared spectra of NGC 1068 along +8\arcdeg, -13\arcdeg, and 
-78.9\arcdeg slits.  The continuum is fit with a silicate extinction curve 
shown as a dashed line.}
\label{n1068abc}
\end{figure}

The mid-infrared spectra of NGC 1068 were obtained at 3 different slit 
position angles, slit PA = +8$^\circ$, +78.9$^\circ$, and -13$^\circ$.  
The spectra along +8\arcdeg and -13\arcdeg slits were spatially resolved along 
the N-S elongated regions while the spectrum along +78.9\arcdeg slit was not 
extended.  We present diffraction-limited total integrated fluxes along each 
slit position in Figure \ref{n1068abc}.  The observed data are shown in a solid 
line while the fit with the silicate extinction profile as a dashed line in each 
figure.  For the spectra along +8\arcdeg and -13\arcdeg slits, we also display 
the relative flux from the central 0.25\arcsec $\times$ 0.25\arcsec rectangular 
region (Central Engine, CE) in the same figures for comparison.  In each slit, 
the flux from CE accounts for a little more than a quarter of the total flux.  As 
mentioned in the previous section, the spectrum along the 
+78.9\arcdeg slit is noisier than the other two spectra due to a notable tracking 
problem from using a faint guide star.  The two spectra along the N-S elongated 
regions look very similar to each other.  They both have similar slopes and show 
significant silicate absorption (Figures 2a and 2b).  But the spectrum 
perpendicular to the N-S elongated regions looks steeper and has less silicate 
absorption than the others (Figure 2c).  

Overall the mid-infrared spectra of NGC 1068 are dominated by the broad silicate 
absorption line without strong emission features.  Although not strong, [S IV] 
emission lines are apparent in all 3 spectra.  We found that the silicate 
absorption lines are fit better with a small shift of the silicate absorption 
center.  Given that the spectral resolution of our data is 0.11 $\micron$, the 
shift is marginal ($\Delta\lambda$ $\lesssim$ 2 $\sigma$ ).  The shifts may, 
however, be real as they all occur toward the shorter wavelength (see Table 2).  
\citet{stu00} have reported a more significant shift in the broad silicate 
absorption feature centered at 9.4 $\micron$ in their ISO-SWS spectra of NGC~1068.  
But it was suspected that the contamination by the large ISO beam (various 
apertures between 14\arcsec $\times$ 20\arcsec and 20\arcsec 
$\times$ 33\arcsec) might have caused the apparent blue shift of the silicate 
absorption center \citep{boc00}.

\begin{figure}
\figurenum{3}
\epsscale{.80}
\plotone{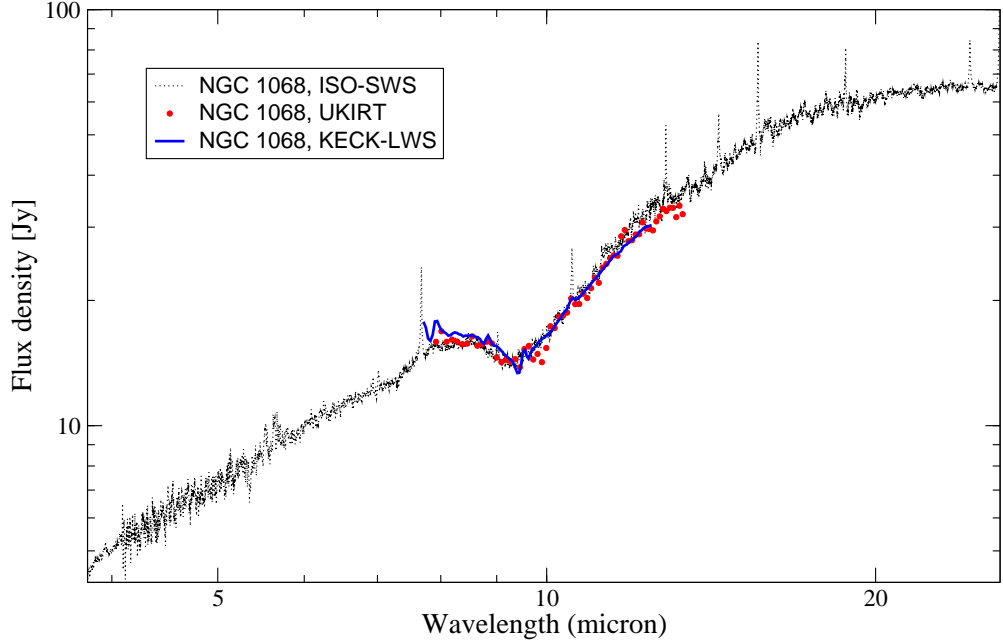}
\caption{The mid-IR spectra of NGC 1068.  The Keck LWS mid-infrared spectrum 
of NGC 1068 was created by averaging two spectra along +8\arcdeg and -13\arcdeg 
slits and scaling to match the ISO-SWS spectrum.  The UKIRT spectrum
is a reproduction of the mid-infrared spectrum of NGC 1068 in Figure 1 from 
\citet{roc91}.  $\lambda$F$\lambda$ was first converted to F$\nu$ and then 
scaled to match the ISO-SWS spectrum.  Dr. Roche kindly provided his UKIRT data 
for this Figure.  The LWS spectrum (0.25\arcsec 
$\times$ 2\arcsec) matches ISO-SWS spectrum (12\arcsec $\times$ 20\arcsec for 
short wavelength and 20\arcsec $\times$ 33\arcsec for long wavelength) and 
UKIRT spectrum (5\arcsec circular aperture) quite well, especially the depth of 
the silicate absortion.  A better spatial resolution of NGC 1068 resulted from 
its proximity may explain that the AGN dominates the mid-IR flux of 
NGC 1068 in these spectra.  D. Lutz kindly provided the ISO-SWS data that 
were used for Figure 1 in \citet{lut00}}
\label{n1068iso}
\end{figure}

In Figure \ref{n1068iso}, we display our mid-IR spectrum (Keck LWS) and the 
UKIRT spectrum \citep{roc91} on top of the ISO-SWS spectrum \citep{lut00} of 
NGC 1068.  For our spectrum in Figure \ref{n1068iso}, two spectra along 
+8\arcdeg and -13\arcdeg slits were averaged together and scaled to match 
the flux of the ISO-SWS spectrum.  The UKIRT spectrum is a reproduction of 
the mid-infrared spectrum of NGC 1068 in Figure 1 from \citet{roc91}.  
$\lambda$F$\lambda$ was first converted to F$\nu$ and then scaled to match the 
ISO-SWS spectrum.\footnote{The actual data from \citet{roc91} has $\sim$15\% 
less flux than what is shown in Figure 1.  The additional flux in ISO-SWS spectrum 
may come from its larger aperture.  The overall shape and the strength of the 
silicate feature of both spectra, however, agree with each other well.}  
Our spectrum matches both the UKIRT spectrum and the ISO-SWS spectrum quite 
well in both the depth and the slope of the silicate absorption feature 
although each spectrum depicts a very different physical scale.  
Our Keck LWS spectrum represents central 0.25\arcsec $\times$ 2\arcsec (
18 pc $\times$ 140 pc in physical scale) while the UKIRT spectrum serves 
about 3 times and the ISO-SWS spectrum about 10 times bigger areas in NGC 1068.  
In general, the mid-IR spectra of Seyfert 2 galaxies obtained with 
a large aperture show very different features than those with a small aperture.  
For example, an average ISO spectrum of 27 Seyfert 2 galaxies shows that strong 
PAH emission lines dominate the mid-IR regions \citep{cla00} due to 
contamination from host galaxy light.  But the mid-IR spectra taken with small 
apertures display a significant silicate absorption feature without strong 
PAH emission lines (\citealt{sie04,soi02}, see \citealt{rhe05} for further 
discussion).  A better spatial resolution of NGC 1068 resulted from its 
proximity may explain why AGN in NGC 1068 dominates both the ISO-SWS and 
other small aperture spectra revealing a significant silicate absorption 
without strong PAH emission lines.  In their ISO-CAM image of NGC 1068 
(Figure 2b), \citet{lef01} shows that in NGC 1068 PAH feature at 7.7$\micron$ 
is very week within $\sim$10\arcsec circular radius from the nucleus but 
gets stronger further out and peaks around 15\arcsec away from the nucleus.  
If NGC 1068 were located at $\sim$20 parsec from the Earth as other typical 
Seyfert 2's, the ISO-SWS spectrum would have easily included strong PAH 
features.

Our mid-IR spectrum (Keck LWS) as well as the UKIRT spectrum and the ISO-SWS 
spectrum shows, however, a stronger silicate absorption than ISO CAM-CVF 
spectrum of the nucleus of NGC 1068 does (see Figure 4 in \citealt{lef01}).  
Applying the same screen model from \citet{dud97}, we found Av $\sim$10 mag 
while \citet{lef01} reported Av $\sim$7 mag.  It is not 
clear why ISO CAM-CVF spectrum shows a flatter and less significant silicate 
absorption in the nucleus of NGC 1068.  \citet{rie75} has also reported a 
weak silicate absorption in NGC 1068.  Their infrared spectrum of NGC 1068 was
not, however, acquired from a spectroscopic observation but inferred from 
several narrow-band photometry results.  As discussed further in section 4, 
such a spectrum does not necessarily agree with the spectroscopic data. 

\begin{figure}
\figurenum{4}
\epsscale{1.0}
\plotone{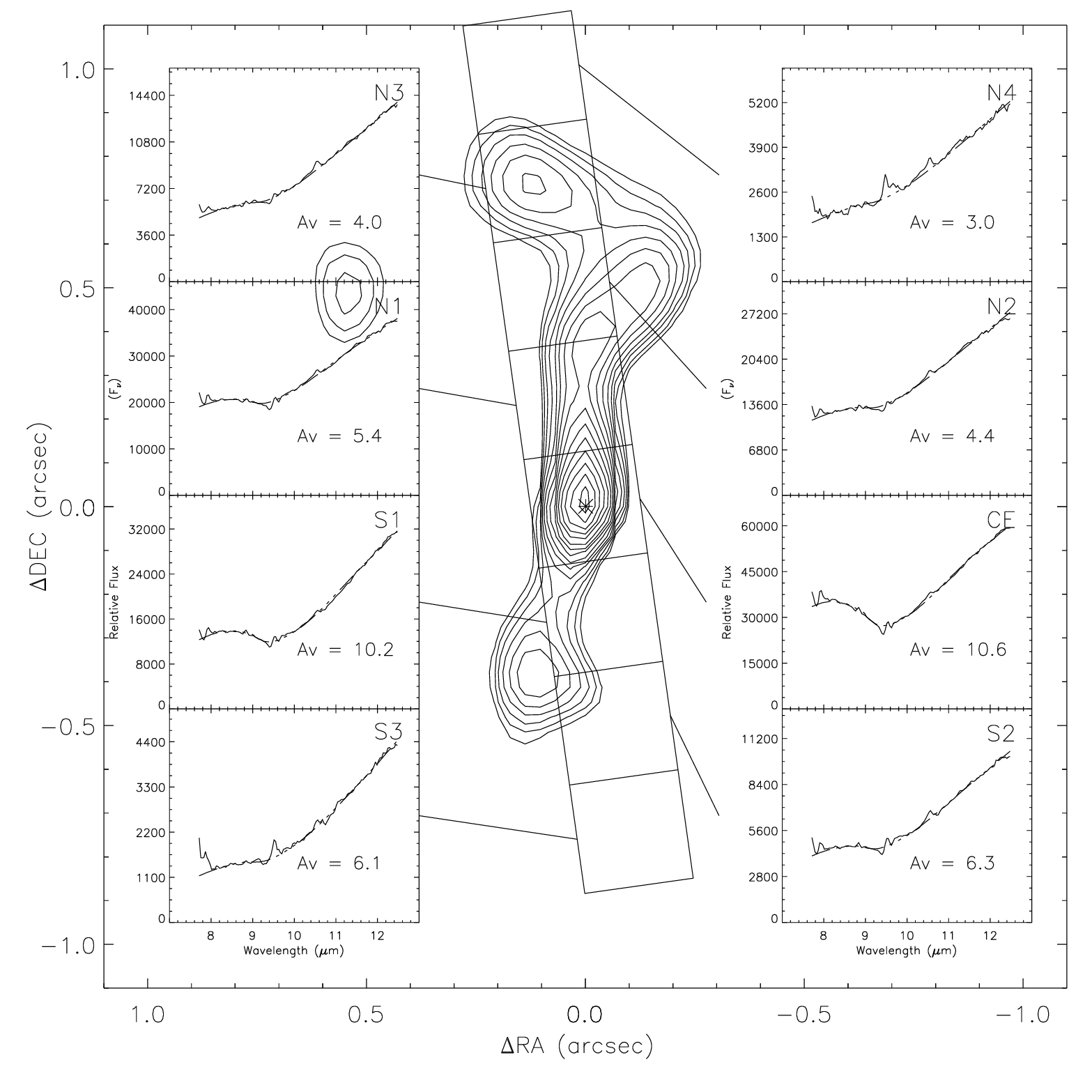}
\caption{Mid-infrared spectra of NGC 1068 along +8\arcdeg slit overlaid on 
12 $\micron$ deconvolved image of NGC 1068 from \citet{boc00}.  Each spectrum 
is produced from a 0.25\arcsec $\times$ 0.25\arcsec square beam and plotted in 
Relative Flux ($F_{\nu}$) as a function of wavelength ($\micron$).  In each 
set, the continuum is fit with a silicate extinction curve shown as a dashed 
line.  }
\label{n1068+8}
\end{figure}

\begin{figure}
\figurenum{5}
\epsscale{1.0}
\plotone{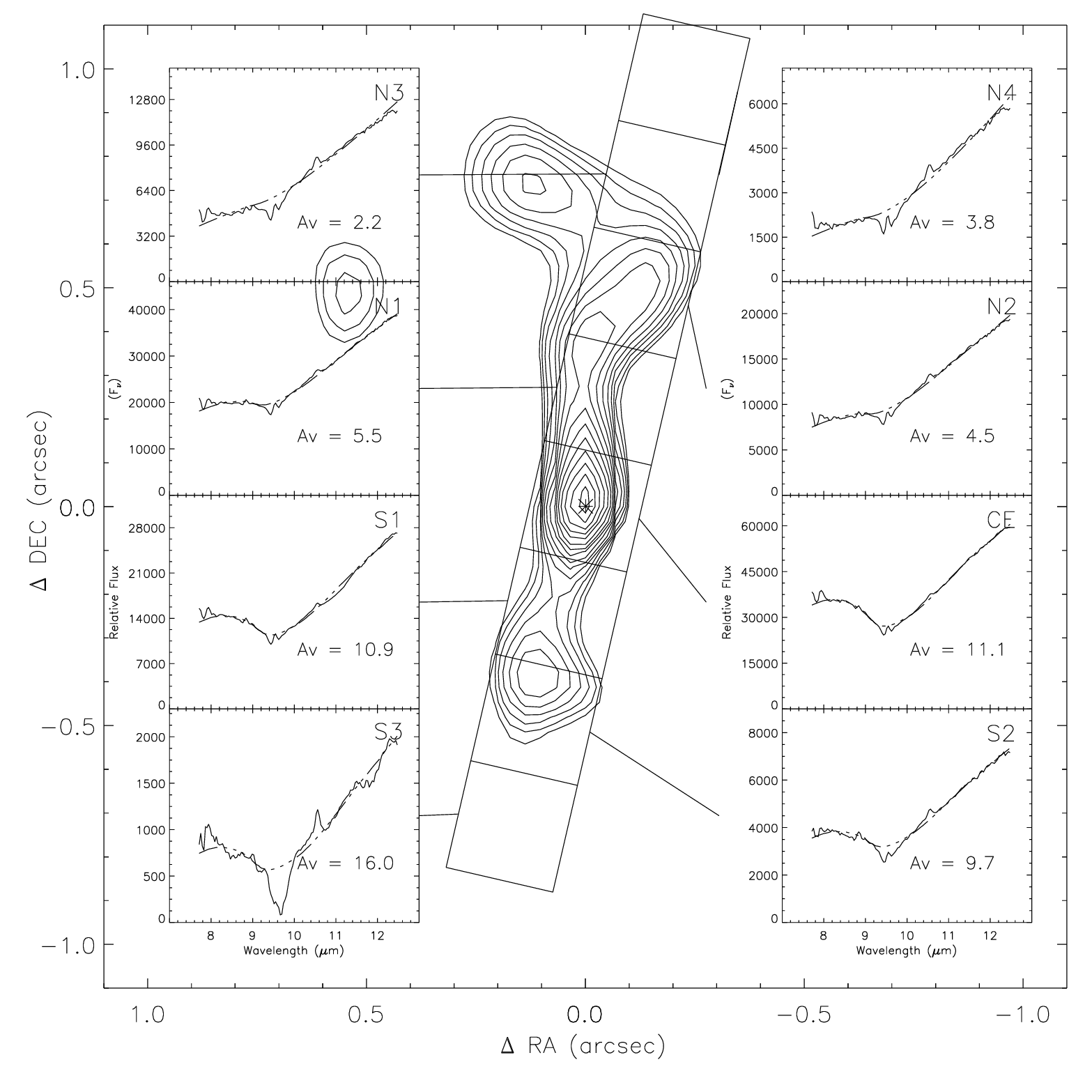}
\caption{Mid-infrared spectra of NGC 1068 along -13\arcdeg slit overlaid on 
12 $\micron$ deconvolved image of NGC 1068 from \citet{boc00}.  Each spectrum 
is produced from a 0.25\arcsec $\times$ 0.25\arcsec square beam and plotted in 
Relative Flux ($F_{\nu}$) as a function of wavelength ($\micron$).  
In each set, the continuum is fit with a silicate extinction curve shown as 
a dashed line.}
\label{n1068-13}
\end{figure}

For the +8\arcdeg and -13\arcdeg slit PA's, we extracted spectra from 8 regions
along the slit: four positions from the northern extended region, one position 
at the nucleus, and three positions from the southern extended region.  Each 
spectrum represents the spectrum in a square beam of 0.25\arcsec $\times$ 
0.25\arcsec.  In Figures \ref{n1068+8} and \ref{n1068-13}, we overlay each 
slit on top of the 12.5 $\micron$ image by \citet{boc00} and present all 8 spectra 
along each slit.  Centered on the nucleus, the slit along +8\arcdeg crosses 
most of the northern extended region, all the way through the upper tongue in 
the north-east tip of the extended image.  In the south, the slit along +8$^\circ$, 
however, misses the bright south-east spot.  The slit along -13\arcdeg covers
the most extended regions well from the southern bright spot to the northern bright 
spot except the upper tongue in the north-east.  Figures \ref{n1068+8} and 
\ref{n1068-13} show that the spectra on the nucleus are bluer but get 
steeper away from the nucleus indicating that the mid-infrared continuum in the 
nucleus is dominated by hotter dust than dust off the nucleus.  The [S IV] 
emission line appears in all the extended regions and the relative strength 
increases with the distance away from the central engine.

The strength of the silicate absorption varies over the extended regions.  It 
is strongest on the nucleus and declines to the north.  This trend shows that 
our line of sight transmits through the largest amount of the cold dust toward 
the nucleus and intercepts less cold dust off from the nucleus.  In addition, 
the silicate absorption remains relatively strong in the southern extended region 
but declines fast in the northern extended region as noted by \citet{boc00}.  
This asymmetry may result from the northern opening of the torus being slightly 
tilted toward our line of sight.  No silicate emission was detected in any 
region, contrasting previous studies which had hinted at silicate line emission 
in the northern extended regions \citep{boc00} or in the southern extended 
regions \citep{gal03}.  This is discussed further in section 4.

In the appendix, we display individual spectra along each slit 
position in Figures \ref{n1068N1} thru \ref{n1068S3} with brief explanations in 
the text.  Each observed spectrum is represented by a solid line along 
with the best silicate extinction fit shown as a dashed line.  In general, the 
fits are reasonable ($\chi^{2}$/N $\lesssim$ 2.0) except in the spectral regions 
contaminated by strong atmospheric absorption lines (i.e. $\lesssim$ 8~$\micron$ 
and 9.2~$\micron$ $\sim$ 9.7~\micron).  Table \ref{tbl-2} summarizes parameters 
used to fit each individual spectrum.  As discussd further in Sect. 4, the 
bestfit parameters, especially $A_{v}$, should be taken with a caution as the
cold screen model may not apply to all the regions, especially in the north.


\section{Discussion}
Previous imaging studies have produced local mid-infrared SEDs and estimates 
of the silicate absorption by combining spatially resolved narrow-band fluxes.  
Both \citet{boc00} and \citet{gal03} provide such SEDs which don't agree with 
each other.  In the northern extended regions, \citet{boc00} show no silicate 
absorption but a smooth continuum, perhaps even silicate in emission.  On 
the other hand, \citet{gal03} find the silicate feature in absorption with a 
depth that increases with the distance from the central engine (CE).  Our 
results do not, however, agree with either of the results above.  Our spectra 
show that the silicate feature appears weak in absorption in the northern 
extended regions and with its depth decreasing with distance from the CE 
instead of increasing suggested by \citet{gal03}.  In the southern extended 
regions, our spectra show significant silicate absorption in the southern 
extended regions consistent with that of \citet{boc00}.  

We believe such discrepancies arise because their local SED were synthesized 
from individual narrow-band images observed at different times and may suffer 
variable seeing conditions.  Under different seeing conditions, it is hard to 
compare accurately the fluxes of different bands.  Our spectra do not,
however, suffer from such uncertainty as all 8 positions within an
individual N-band spectra were obtained similtaneously.

The presence of the prominent structure to the north in the mid-infrared image 
was used to indicate that the north axis of the torus opening is tilted towards 
our line of sight \citep{boc00}.  The presence of strong silicate absorption 
towards the southern extended region supports this picture as it indicates 
obscuration from optically thick dust grains in the south.  However, the 
presence of weak silicate absorption to the north is interesting.  If we assume 
a simple torus model in which the northern openning is oriented towards us, 
then we might expect to see hot dust on the inner edge in emission asssuming 
a similar dust composition as the obscuring material.  We may not, however, 
have the same kind of cold foreground screen toward the hot diffuse dust in 
the north as to the nucleus and the south; thus, it is difficult to draw 
any firm conclusion by applying the same model to all regions and comparing the 
strength of the silicate features quatitatively based on the model fit such 
as the visual extinctions, $A_{v}$.  A radiative transfer model is required to 
address properly the apparent differences in the strength of the silicate 
feature at various locations because dust temperature distribution and 
geometric effect affect the silicate feature amplitude.  For non-spatially 
resolved mid-IR spectra of NGC 1068, many theoretical works have already been 
done using various radiative transfer models over last a couple of decades 
\citep{pie92,pie93,gra94,gra97,row95,nen02}.

\begin{figure}
\figurenum{6}
\plotone{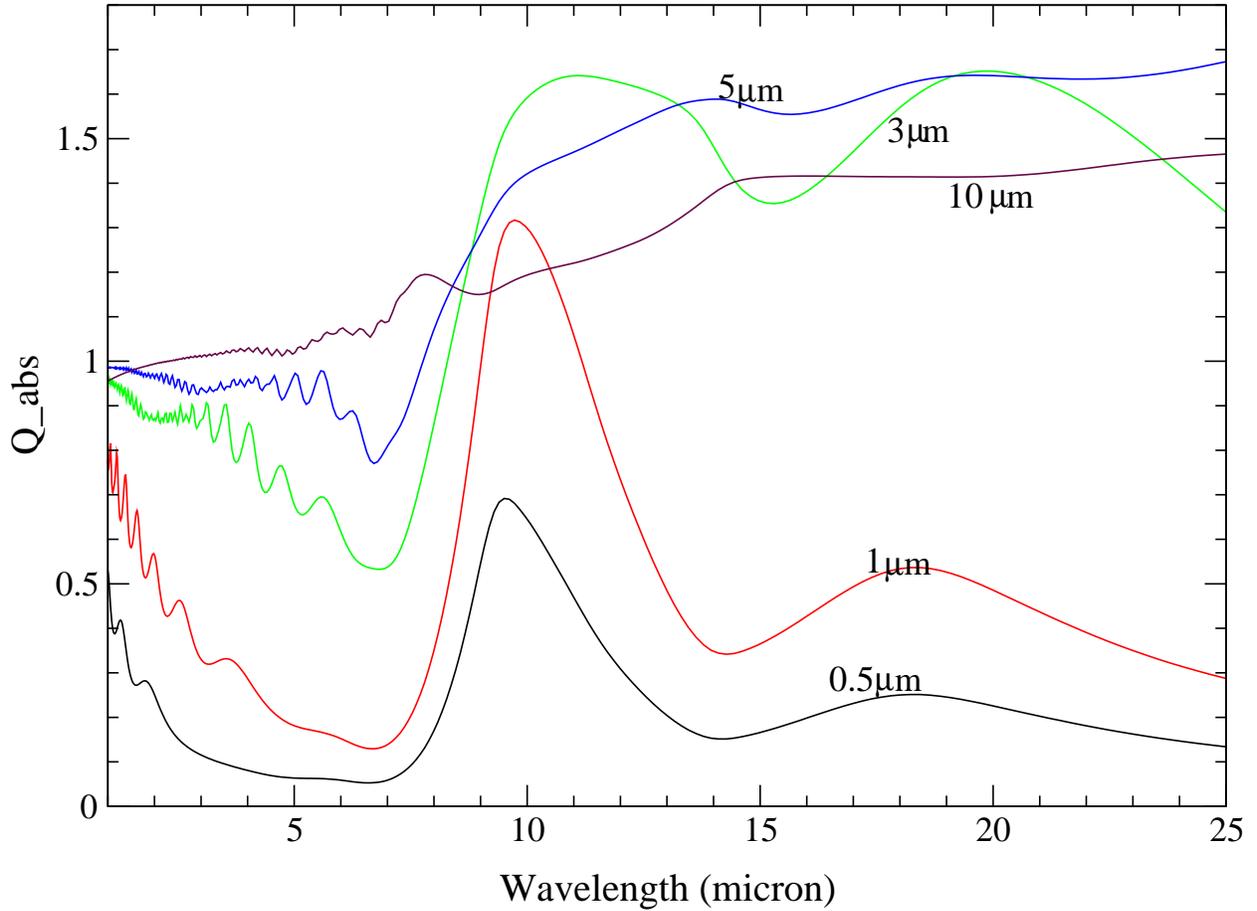}
\caption{Absorption efficiency of silicate grains vs. its size.  This figure
is produced using a code used in Figure 5 in Draine and Lee (1984).  Aigen Li
kindly provided the code for this paper.}
\label{silicate_eff}
\end{figure}

In the absence of a radiative transfer model, one may still speculate a depletion 
of small grains within the elogated extended regions as one of the possibilities 
to explain the observed spectra in the north.  The possible depletion of 
small grains in the AGN environment has been suggested by various authors 
\citep{mai01,wei01,rhe05}.  The fact that the northern extended regions 
reside inside the ionization cone renders further speculation.  The formation 
of the ionization cone in NGC 1068 is discussed in three different ways by 
\citet{boc00}: Relativistic beaming, Dust absorption, and Electron scattering.  
In all three ways, the dust grains inside the ionization cones are directly 
heated by very strong UV radiation from the nuclear engine.  This direct 
heating and other processes in the cone can destroy small grains.  Figure 
\ref{silicate_eff} shows that only small grains less than 3 $\micron$ 
effectively give rise to the silicate feature seen in the mid-infrared spectra 
of Seyfert galaxies.  The effective destruction of small grains by strong 
nuclear radiation may explain the observed lack of silicate emission from the 
northern extended regions.  The weak silicate absorption in the 
northern extended regions could be created due to absorption by cold dust 
grains puffed up from the torus between the observer and the hot dust grains 
that emit the continuum.

\begin{figure}
\figurenum{7}
\plotone{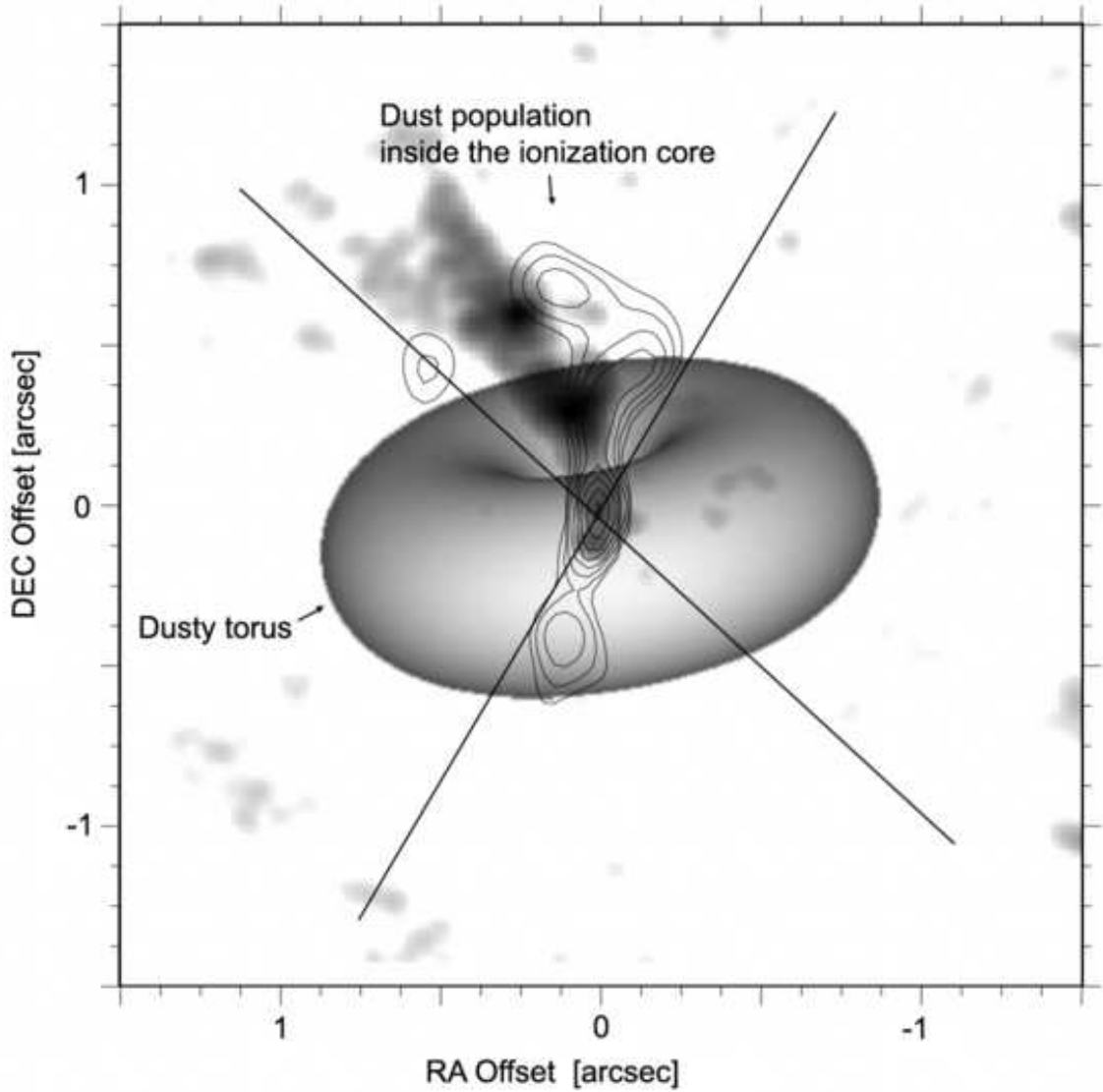}
\caption{Dual dust populations in the nuclear environment of NGC 1068:
one inside the ionization cone and the other in the form of a torus.  The small 
grains are depleted in the dust clouds inside the ionization cone giving rise 
to nearly flat mid-ir spectra in the northern extended regions.  But these 
grains are present in the torus and produce strong 9.7 $\micron$ silicate 
absorption.  The background image of NGC 1068 was taken from Figure 4 in 
Bock et al.\ (2000) in which the contour plot of the 12.5 $\micron$ image was 
superimposed on the 5 GHz map of \citet{gal96}.  The size and shape of the dusty 
torus in the figure does not necessarily depict the true torus as this could not 
be determined by our data.}
\label{ddp}
\end{figure}

In contrast, the small grains in the cold outer regions of the dusty torus 
are shielded from and therefore survive the intense UV radiation.  These 
small silicate grains can effectively absorb the continuum and create the
silicate absorption feature observed in the mid-infrared spectra of NGC 1068. 
Our spectra show that silicate absorption remains strong even to the location 
South 3.  If the silicate grains responsible for the apparent silicate 
absorption belong to the dusty torus, this indicates that the dusty torus 
should extend at least tens of parsecs in vertical direction.  A recent 
interferometric mid-infrared observation of NGC 1068, however, has revealed 
a parsec scale structure believed to be the dusty torus \citep{jaf04}.  
Perhaps the outer cold regions of a dusty torus do not radiate significantly 
at 10 $\micron$ because they are too cold.  The parsec structure found in 
the interferometric image most likely come from the heated inner regions of 
the dusty torus because the interferometric mid-infrared spectra indicate the
presence of cold small grains in the foreground by showing deep silicate 
absorption.  It is perhaps that the outer radius of the torus reaches out 
to more than tens of parsec.  The dust distribution in the nuclear environment 
of NGC 1068 is summarized in Figure \ref{ddp}.

\section{Summary and Conclusions}

We used the Long Wavelength Spectrometer (LWS) at Keck I telescope with a 
0.25\arcsec slit to obtain spatially-resolved spectra of NGC 1068 at three 
different slit positions, slit PA = +8$^\circ$, -13$^\circ$, and +78.9$^\circ$. 
Overall our integrated spectra agrees with the previous mid-IR spectra well.
In NGC 1068, our resolved spectra showed that the silicate absorption was 
strongest on the nucleus (Central Engine) indicating that our line of sight 
transmited through the largest amount of cold obscuring dust materials to the 
nucleus.  Furthermore, the strength of the silicate absorption declines fast 
to the north but remains relatively strong to the south.  This asymmetry in 
the strength of the silicate absorption as well as the larger extended 
emission regions to the north suggests that the north pole of the torus 
opening is inclined towards us and that the southern extended regions are 
behind the obscuring dusty torus.  Based on the lack of strong silicate 
emission from the north, we considered the dual dust population model for 
AGN in which two different dust populations exist in the ionization cones 
and in the dusty torus.  The dust population in the ionization cones may lack 
small silicate grains or small grains in general and give rise to 
featureless continuum to the north while the dust population in the dusty 
torus contains plenty of small silicate grains and causes the apparent deep 
silicate absorption lines to the nucleus and the south in NGC 1068.  A 
detailed analysis using a radiative transfer model would provide a better 
understanding to our spatially resolved mid-IR spectra.

\acknowledgments
We like to thank Randy Campbell for support during the observation with LWS at 
Keck I telescope.  We thank Chris Dudley at Naval Observatory for his valuable
comments as well as the cold screen model of the 9.7 $\micron$ silicate absorption 
feature used in this paper.  We appreciate Dieter Lutz and Pat Roche for their 
ISO-SWS and UKIRT spectra reproduced Figure \ref{n1068iso}.  Jamie J. Bock also 
kindly provided the images of NGC 1068 reproduced in Figures \ref{n1068+8} 
and \ref{n1068-13}.  We are grateful to Aigen Li for providing the code 
to produce Figure \ref{silicate_eff}.  We acknowledge useful conversations with 
Ari Laor, Matt Malkan and Eric Becklin. J.R. thanks K. Cha and I. Song for their 
help in creating Figure \ref{ddp}.

\appendix

\section{Appendix material}

\subsection{Mid-IR spectra at the Central Engine}

\begin{figure}
\figurenum{8}
\epsscale{0.8}
\plotone{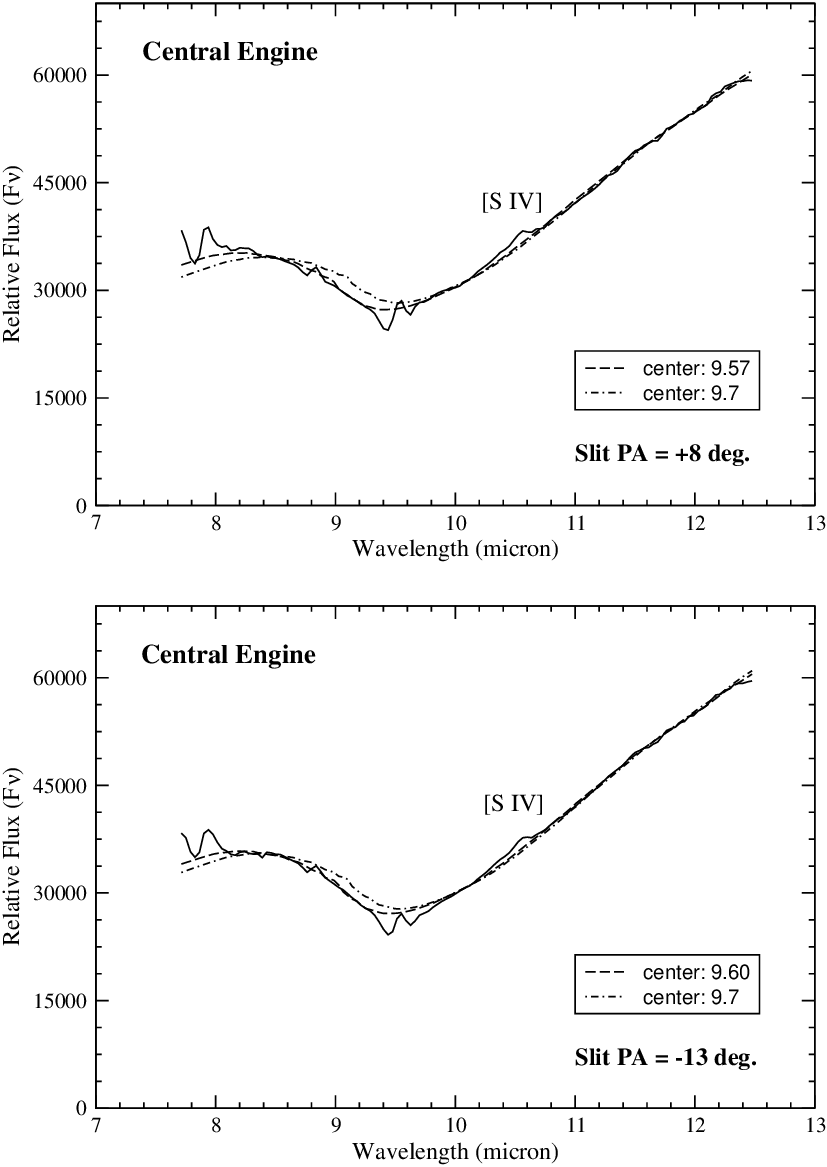}
\caption{Central Engine: Mid-infrared spectra of NGC 1068.  The top panel is from
the slit along +8\arcdeg and the bottom panel from -13\arcdeg.  Each spectrum 
is produced from a 0.25\arcsec $\times$ 0.25\arcsec square beam.  In each 
set, the observed spectrum is fit with the silicate extinction curve.  A dashed line
represents the best fit and a dashed-dotted line the fit with a center at 9.7 \micron.}
\label{n1068CE}
\end{figure}

Figure \ref{n1068CE} depicts the nuclear spectra of NGC 1068 at the central
0.25\arcsec $\times$ 0.25\arcsec location (Central Engine, CE) along the +8\arcdeg 
slit in the top panel and the -13\arcdeg slit in the bottom panel.  Produced 
essentially from the same region, they look almost identical except at different 
relative flux levels.  The spectra at the central engine is the strongest 
overall and contain at least one-quarter of the total flux in the slit (see Figure 
\ref{n1068abc}).  The spectra at the central engine rises at the 
short wavelength end indicating that the nuclear flux at the mid-infrared is 
dominated by hotter dust than dust off the nucleus.  The silicate
absorption is also strongest here implying the 
line of sight transmits through the largest amount of cold silicate grains.
The silicate absorption line is best fit with the silicate extinction profile 
whose centers fall on 9.57 $\micron$ in the top panel and 9.60 $\micron$ in the 
bottom panel instead of 9.7 \micron.  A weak [S IV] emission line appears at 
10.5 $\micron$ in both spectra. 

\subsection{Mid-IR spectra at North 1}

\begin{figure}
\figurenum{9}
\epsscale{0.8}
\plotone{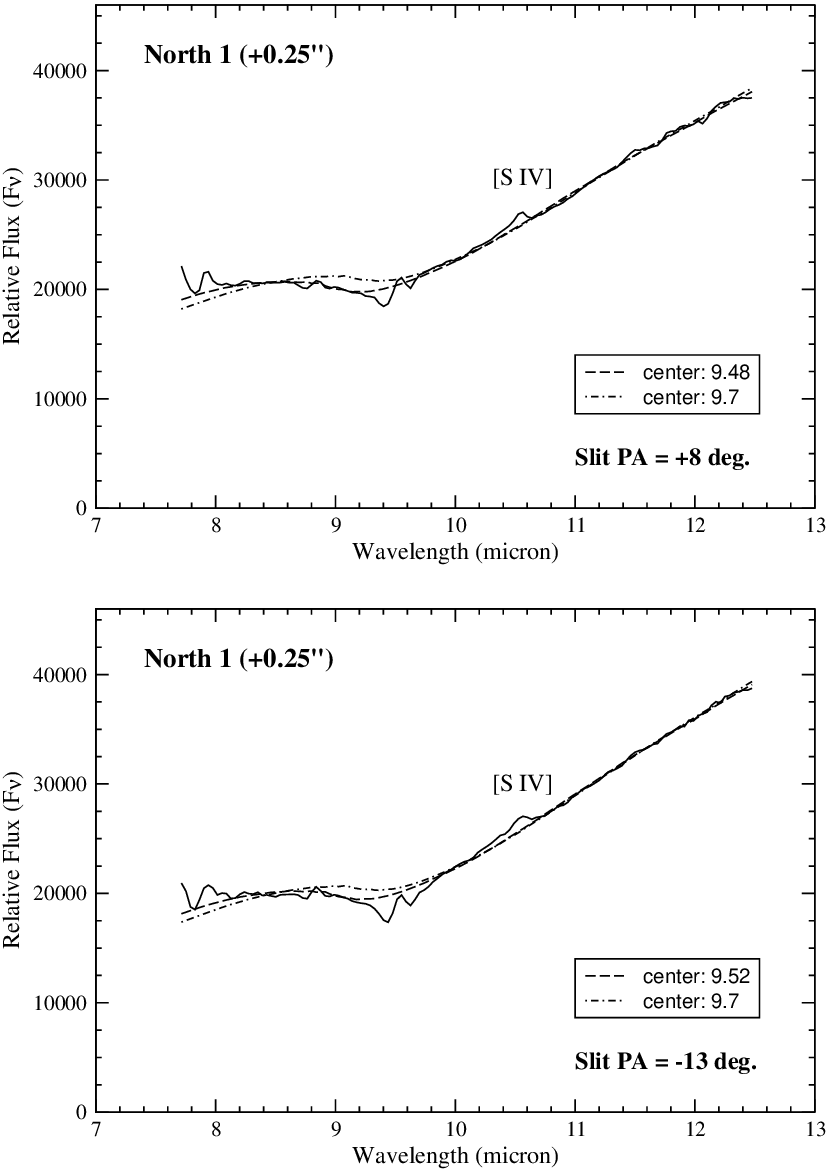}
\caption{Norh 1: Mid-infrared spectra of NGC 1068.  The top panel is from
the slit along +8\arcdeg and the bottom panel from -13\arcdeg.  Each spectrum 
is produced from a 0.25\arcsec $\times$ 0.25\arcsec square beam.  In each 
set, the observed spectrum is fit with the silicate extinction curve.  A dashed line
represents the best fit and a dashed-dotted line the fit with a center at 9.7 \micron.}
\label{n1068N1}
\end{figure}

Figure \ref{n1068N1}, North 1, shows the spectra of NGC 1068 at 0.25\arcsec off 
the central peak to the north along the +8\arcdeg slit in the top panel and the 
-13\arcdeg slit in the bottom panel.  Overall they look similar to each other.  
Flux decreases significantly to about one half of the CE and the spectra are 
much steeper.  The silicate absorption is still significant but greatly reduced 
compared to the CE and even to South 1.  Figure \ref{n1068N1} shows that the 
silicate extinction profile fits the observed data extremely well at both slit 
positions.  The best fit centers of the silicate absorption lines are to the blue
of the nominal wavelength in both panels.  The [S IV] emission lines have a larger
equivalent width than at the CE and appear at 10.5 $\micron$ in both spectra. 

\subsection{Mid-IR spectra at North 2}

\begin{figure}
\figurenum{10}
\epsscale{0.8}
\plotone{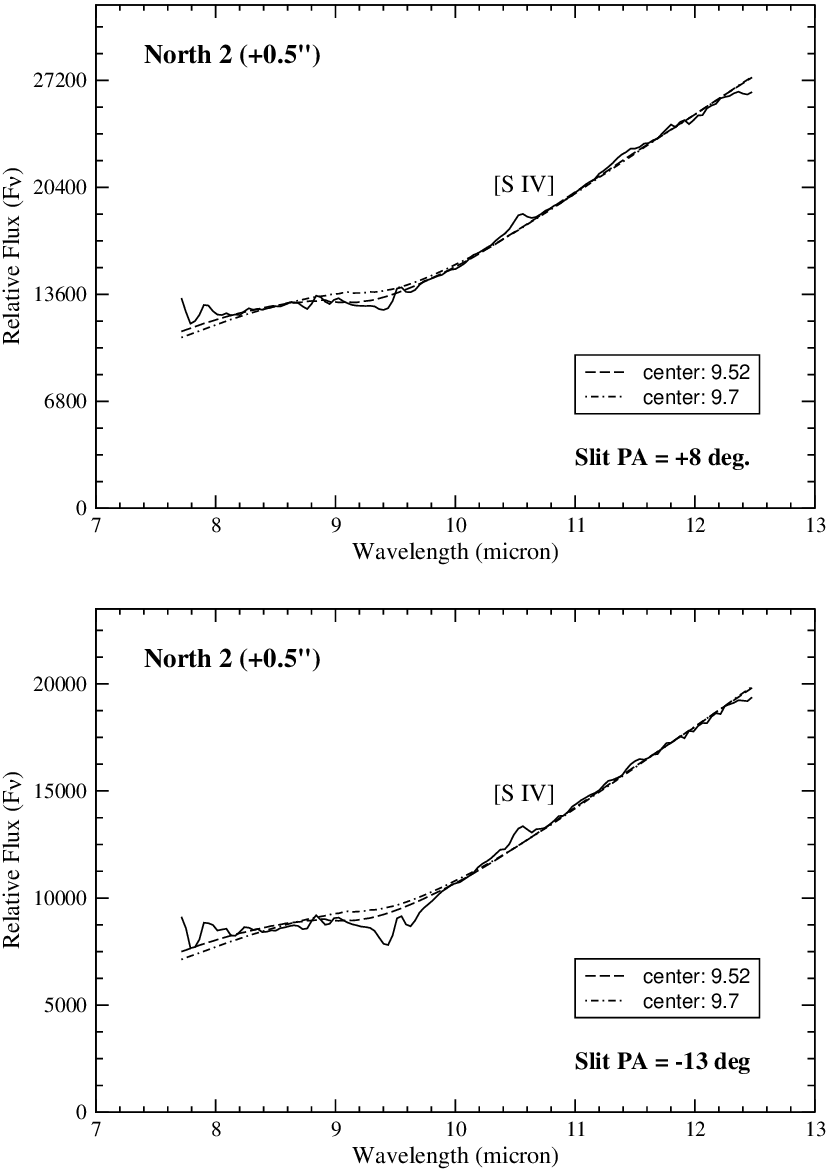}
\caption{North 2: Mid-infrared spectra of NGC 1068.  The top panel is from
the slit along +8\arcdeg and the bottom panel from -13\arcdeg.  Each spectrum 
is produced from a 0.25\arcsec $\times$ 0.25\arcsec square beam.  In each 
set, the observed spectrum is fit with the silicate extinction curve.  A dashed line
represents the best fit and a dashed-dotted line the fit with a center at 9.7 \micron.}
\label{n1068N2}
\end{figure}

Figure \ref{n1068N2}, North 2, shows the spectra of NGC 1068 at 0.50\arcsec north 
of the central peak along the +8\arcdeg slit in the top panel and the -13\arcdeg 
slit in the bottom panel.  In both panels, the spectra appear almost flat with 
weak silicate absorption in the middle.  The flux drops to around 30$\%$ 
of the CE.  The silicate extinction curve fit continues to match the observed data 
well with its center shifted to 9.54 $\micron$ in both the top and bottom panels.   

\subsection{Mid-IR spectra at North 3}

\begin{figure}
\figurenum{11}
\epsscale{0.8}
\plotone{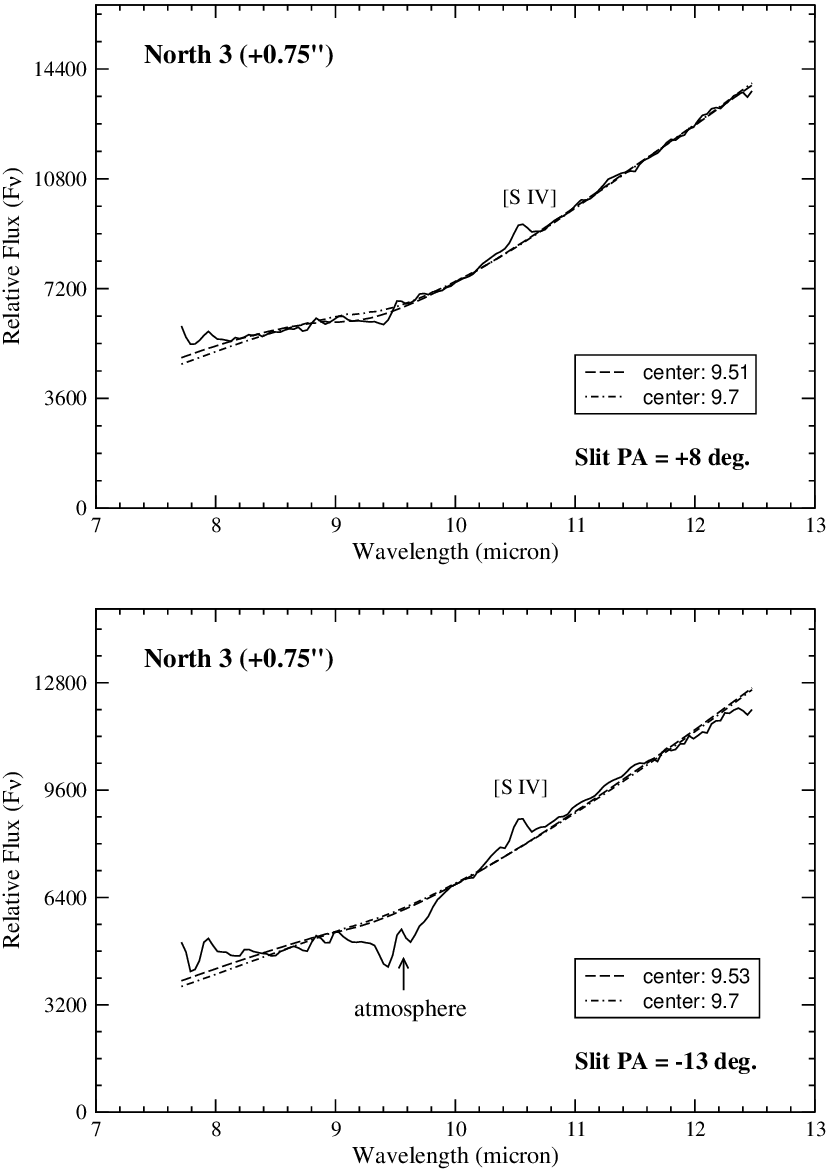}
\caption{North 3: Mid-infrared spectra of NGC 1068.  The top panel is from
the slit along +8\arcdeg and the bottom panel from -13\arcdeg.  Each spectrum 
is produced from a 0.25\arcsec $\times$ 0.25\arcsec square beam.  In each 
set, the observed spectrum is fit with the silicate extinction curve.  A dashed line
represents the best fit and a dashed-dotted line the fit with a center at 9.7 \micron.}
\label{n1068N3}
\end{figure}

Figure \ref{n1068N3}, North 3, shows the spectra of NGC 1068 at 0.75\arcsec off 
the central peak to the north along the +8\arcdeg slit in the top panel and the 
-13\arcdeg slit in the bottom panel.  The spectrum in the top panel (from the 
+8\arcdeg slit) is approximately centered on the 12 $\micron$ feature called 
the ``tongue'', which is to the North-north-east of the nucleus.  In both spectra,
the flux is around 5$\%$ of the CE.  The silicate absorption continues 
to appear weak while the relative [S IV] emission is stronger than in 
North 2.

\subsection{Mid-IR spectra at North 4}

\begin{figure}
\figurenum{12}
\epsscale{0.8}
\plotone{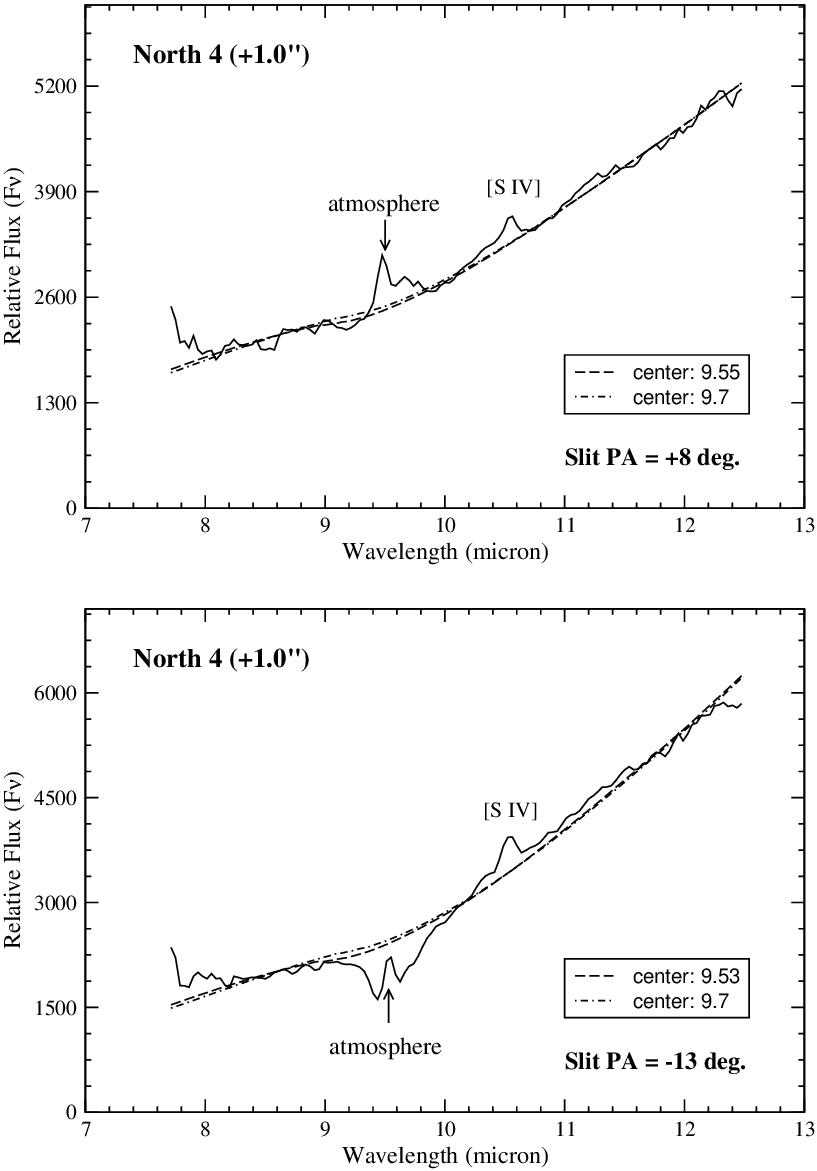}
\caption{North 4: Mid-infrared spectra of NGC 1068.  The top panel is from
the slit along +8\arcdeg and the bottom panel from -13\arcdeg.  Each spectrum 
is produced from a 0.25\arcsec $\times$ 0.25\arcsec square beam.  In each 
set, the observed spectrum is fit with the silicate extinction curve.  A dashed line
represents the best fit and a dashed-dotted line the fit with a center at 9.7 \micron.}
\label{n1068N4}
\end{figure}

Figure \ref{n1068N4}, North 4, depicts the spectra of NGC 1068 at 1.0\arcsec off 
the central peak to the north along the +8\arcdeg slit in the top panel and the 
-13\arcdeg slit in the bottom panel.  Both spectra have low fluxes and appear 
noisy as they represent the regions outside the contours on the 12 $\micron$ map.  
The spectra rise toward the long wavelength.  The [S IV] equivalent width is 
stronger than at any other region except South 3 along the -13\arcdeg slit.

\subsection{Mid-IR spectra at South 1}

\begin{figure}
\figurenum{13}
\epsscale{0.8}
\plotone{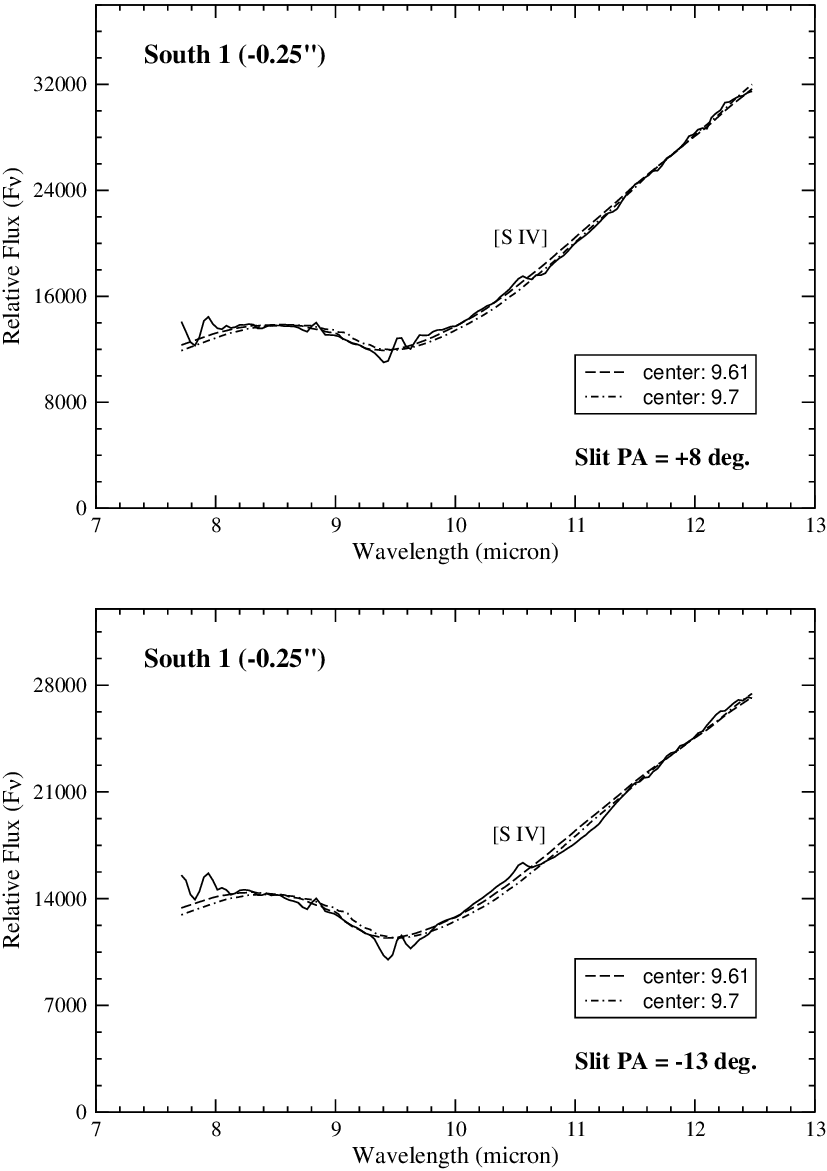}
\caption{South 1: Mid-infrared spectra of NGC 1068.  The top panel is from
the slit along +8\arcdeg and the bottom panel from -13\arcdeg.  Each spectrum 
is produced from a 0.25\arcsec $\times$ 0.25\arcsec square beam.  In each 
set, the observed spectrum is fit with the silicate extinction curve.  A dashed line
represents the best fit and a dashed-dotted line the fit with a center at 9.7 \micron.}
\label{n1068S1}
\end{figure}

Figure \ref{n1068S1}, South 1, depicts the spectra of NGC 1068 at 0.25\arcsec 
south of the central peak along the +8\arcdeg slit in the top panel and the 
-13\arcdeg slit in the bottom panel.  Even with a substantial decline in flux, 
the silicate absorption at South 1 remains as strong as in the CE contrary to 
North 1 in which the silicate absorption drops significantly, indicating that a 
large amount of obscuring cold dust still remains in the foreground.  Here the 
amount of the shift in the centers of the silicate absorption lines in both 
panels are least among all the regions.  The [S IV] equivalent width appears as 
weak as at the CE.

\subsection{Mid-IR spectra at South 2}

\begin{figure}
\figurenum{14}
\epsscale{0.8}
\plotone{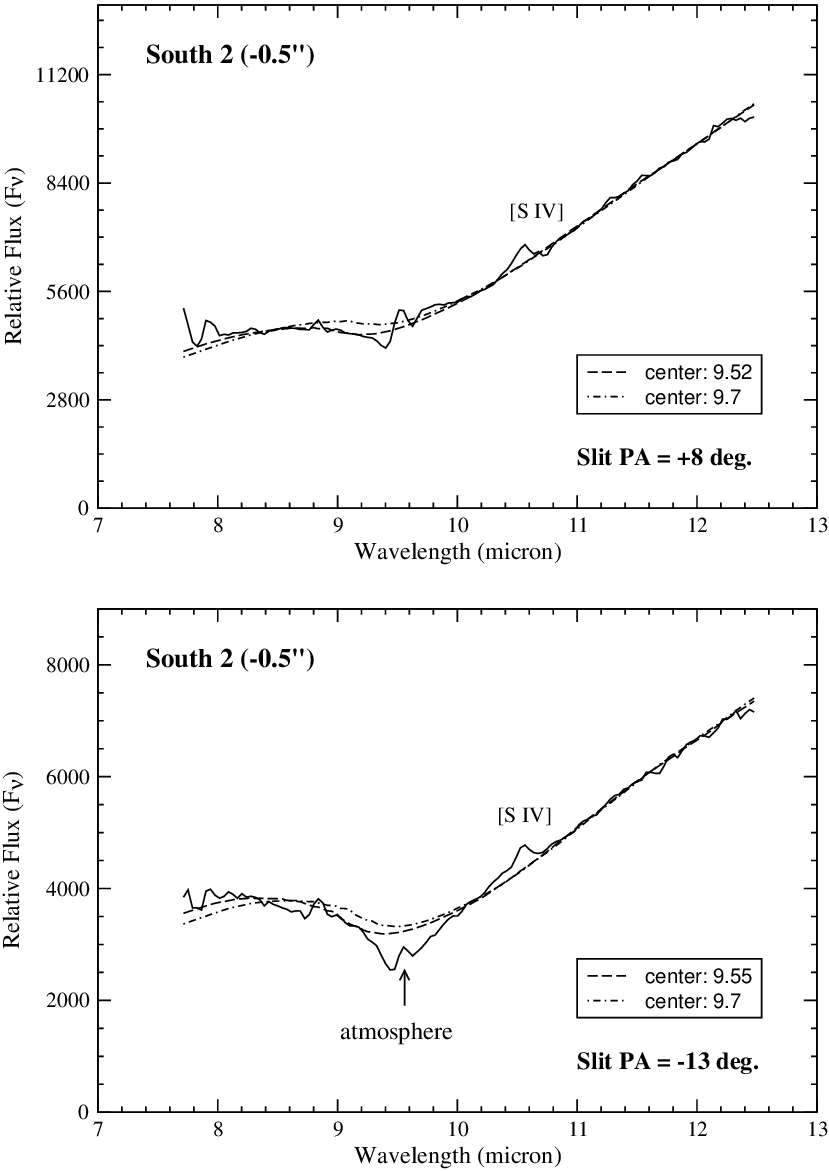}
\caption{South 2: Mid-infrared spectra of NGC 1068.  The top panel is from
the slit along +8\arcdeg and the bottom panel from -13\arcdeg.  Each spectrum 
is produced from a 0.25\arcsec $\times$ 0.25\arcsec square beam.  In each 
set, the observed spectrum is fit with the silicate extinction curve.  A dashed line
represents the best fit and a dashed-dotted line the fit with a center at 9.7 \micron.}
\label{n1068S2}
\end{figure}

Figure \ref{n1068S2}, South 2, depicts the spectra of NGC 1068 at 0.50\arcsec south 
of the central peak along the +8\arcdeg slit in the top panel and the -13\arcdeg 
slit in the bottom panel.  The bottom spectrum presented in South 2 along the 
-13\arcdeg slit is from the southern bright knot in the 12 $\micron$ image but the 
top spectrum along the +8\arcdeg slit is significantly to the west of the knot.  
In the bottom panel, the silicate absorption remains strong and the flux rises in 
the short wavelength end like in CE.  In contrast, the top panel shows that the 
silicate absorption is weak and the spectrum is flat.  The centers of the silicate 
absorption lines appear blue-shifted.  The relative [S IV] emission lines appear 
stronger than both at the CE and South 1 in both panels.

\subsection{Mid-IR spectra at South 3}

\begin{figure}
\figurenum{15}
\epsscale{0.8}
\plotone{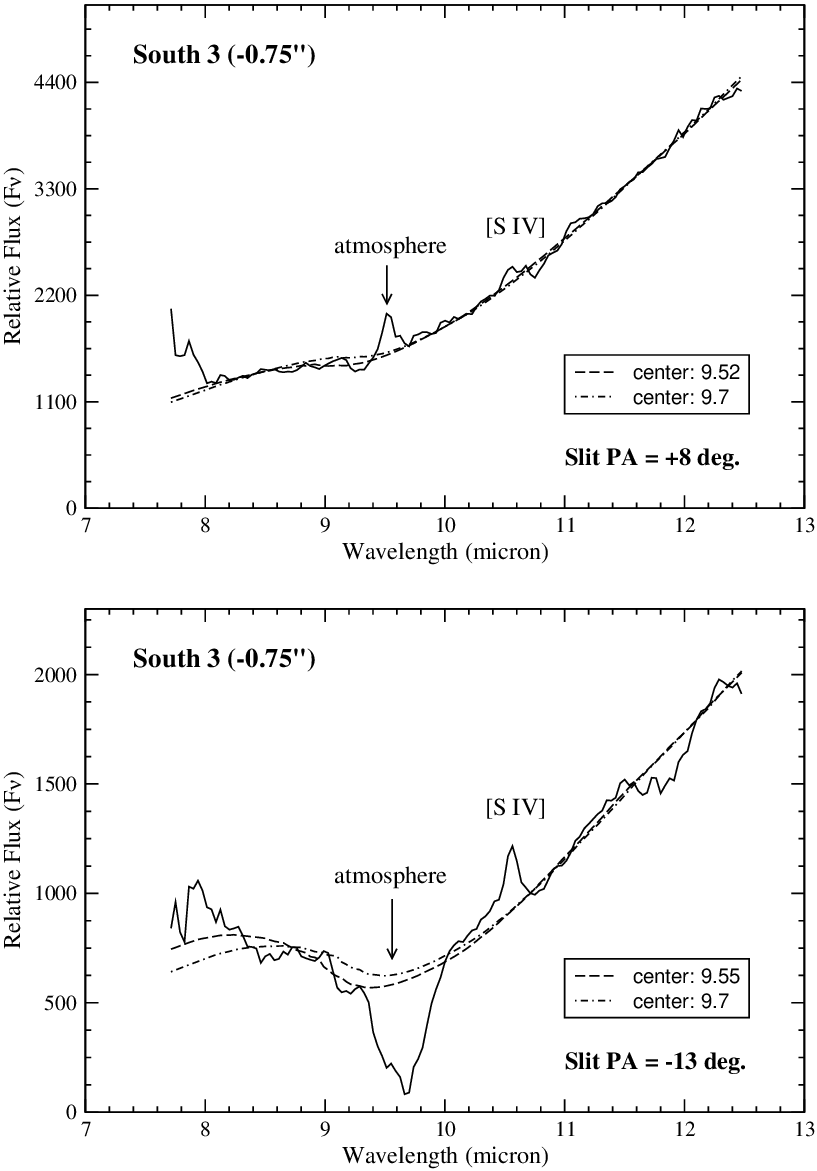}
\caption{South 3: Mid-infrared spectra of NGC 1068.  The top panel is from
the slit along +8\arcdeg and the bottom panel from -13\arcdeg.  Each spectrum 
is produced from a 0.25\arcsec $\times$ 0.25\arcsec square beam.  In each 
set, the observed spectrum is fit with the silicate extinction curve.  A dashed line
represents the best fit and a dashed-dotted line the fit with a center at 9.7 \micron.}
\label{n1068S3}
\end{figure}

Figure \ref{n1068S3}, South 3, depicts the spectra of NGC 1068 at 0.75\arcsec south
of the central peak along the +8\arcdeg slit in the top panel and the -13\arcdeg 
slit in the bottom panel.  This position is below the lowest contour in the 
12 $\micron$ image, and both spectra are noisy and the flux drops to 2$\%$ of the 
CE.  In the top panel, the spectrum is very red and shows little silicate absorption.  In the 
bottom panel, however, the spectrum is flat and displays strong silicate absorption.  The 
relative [S IV] emission lines have higher equivalent width than in South 2.


\clearpage 

\clearpage 

\clearpage 

\clearpage 
\clearpage 

\clearpage 

\begin{deluxetable}{rrrrrrr}
\tablecolumns{7}
\tablewidth{0pc}
\tablecaption{OBSERVATION SUMMARY \label{tbl-1}}
\tablehead{
\colhead{}  & \colhead{Slit}  &\colhead{}  &\colhead{}  &\colhead{}  & 
\colhead{Total Integration\tablenotemark{b}} &\colhead{Date of}\\
\colhead{Object} & \colhead{PA(\arcdeg)} & \colhead{Type}   &  \colhead{cz}   &
\colhead{airmass\tablenotemark{a}}  &\colhead{Time(sec)} & \colhead{Observation} 

}
\startdata
NGC 1068 &+8    &S1.8    &1.003793 & 1.06 &504  & 2003 Sep \\
NGC 1068 &+78.9 &S1.8    &1.003793 & 1.10 &504  & 2003 Sep \\
NGC 1068 &-13   &S1.8    &1.003793 & 1.07 &504  & 2003 Sep \\
\enddata
\tablenotetext{a}{average for all three runtime sets at each slit position.} 
\tablenotetext{b}{for all three runtime sets.  Intergration time for each runtime set
is 168 seconds.} 
\end{deluxetable}
\clearpage 

\begin{deluxetable}{rrrrrr}
\tablecolumns{6}
\tablewidth{0pc}
\tablecaption{SPATIALLY RESOLVED 9.7 $\micron$ SILICATE FEATURE IN NGC~1068\label{tbl-2}}
\tablehead{
\colhead{}  & \colhead{}  &
\multicolumn{4}{c}{Fitting Parameters}\\
\cline{3-6} \\
\colhead{Location} & \colhead{Slit PA} & \colhead{Center(\micron)} 
& \colhead{Index\tablenotemark{a}} & \colhead{$A_{v}$} &
\colhead{$\chi^{2}$/N}
}
\startdata
North 4 &+8\arcdeg   &9.55 &2.35 &2.98  &1.19 \\
        &-13\arcdeg &9.53 &2.95 &3.83  &1.25 \\
North 3 &+8\arcdeg   &9.51 &2.18 &4.01  &1.11 \\
        &-13\arcdeg &9.53 &2.55 &2.17  &1.33 \\
North 2 &+8\arcdeg   &9.52 &1.88 &4.44  &1.05 \\
        &-13\arcdeg &9.52 &2.05 &4.46  &1.31 \\
North 1 &+8\arcdeg   &9.48 &1.47 &5.42  &1.42 \\
        &-13\arcdeg &9.52 &1.63 &5.47  &1.73 \\
Central &+8\arcdeg   &9.57 &1.29 &10.59 &1.16 \\
Engine  &-13\arcdeg &9.60 &1.28 &11.12 &0.98 \\
South 1 &+8\arcdeg   &9.61 &2.05 &10.20 &1.27 \\
        &-13\arcdeg &9.61 &1.56 &10.90 &1.54 \\
South 2 &+8\arcdeg   &9.52 &2.01 &6.34  &0.92 \\
        &-13\arcdeg &9.58 &1.58 &9.72  &1.10 \\
South 3 &+8\arcdeg   &9.52 &2.87 &6.10  &0.81 \\
        &-13\arcdeg &9.55 &2.12 &15.98 &1.80 \\
\enddata
\tablenotetext{a}{The power-law index, $\alpha$, for $F_{\nu}$ $\sim$ $\nu^{-\alpha}$.}
\end{deluxetable}

\end{document}